# Polarization-controlled anisotropic coding metamaterials at terahertz frequencies

Shuo Liu[1,2,*], Tie Jun Cui[1,3,*+], Quan Xu[4], Di Bao[1,2], Liangliang Du[4], Xiang Wan[1,2], Wen Xuan Tang[1,2], Chunmei Ouyang[4], Xiao Yang Zhou[1,2,5], Hao Yuan[5], Hui Feng Ma[1,2], Wei Xiang Jiang[1,2], Jiaguang Han[4], Weili Zhang[3,4], and Qiang Cheng[1,3]

[1] State Key Laboratory of Millimeter Waves, Southeast University, Nanjing 210096, China
[2] Synergetic Innovation Center of Wireless Communication Technology, Southeast University, Nanjing 210096, China
[3] Cooperative Innovation Centre of Terahertz Science, No.4, Section 2, North Jianshe Road, Chengdu 610054, China
[4] Center for Terahertz Waves and College of Precision Instrument and Optoelectronics Engineering, Tianjin University, Tianjin 300072, China
[5] Jiangsu Xuantu Technology Co., Ltd., 12 Mozhou East Road, Nanjing 211111, China

[*] These authors contribute equally to this work.

[+] Corresponding author: E-mail: tjcui@seu.edu.cn.

**Metamaterials based on effective media have achieved a lot of unusual physics (e.g. negative refraction and invisibility cloaking) owing to their abilities to tailor the effective medium parameters that do not exist in nature. Recently, coding metamaterials have been suggested to control electromagnetic waves by designing the coding sequences of digital elements '0' and '1', which possess opposite phase responses. Here, we propose the concept of anisotropic coding metamaterial at terahertz frequencies, in which coding behaviors in different directions are dependent on the polarization status of terahertz waves. We experimentally demonstrate an ultrathin and flexible polarization-controlled anisotropic coding metasurface functioning in the terahertz regime using specially-designed coding elements. By encoding the elements with elaborately-designed digital sequences (in both 1 bit and 2 bits), the *x*- and *y*-polarized reflected waves can be deflected or diffused independently in three dimensions. The simulated far-field scattering patterns as well as near-electric-field distributions are given to illustrate the bifunctional performance of the encoded metasurface, which show good agreement to the measurement results. We further demonstrate the abilities of anisotropic coding metasurface to generate beam splitter and realize anomalous reflection and polarization conversion simultaneously, providing powerful controls of differently-polarized terahertz waves. The proposed method enables versatile beam behaviors under orthogonal polarizations using a single metasurface, and hence will promise interesting terahertz devices.**

It has been shown that artificial metamaterials made of subwavelength structures can be described by effective media with continuous medium parameters[1-3]. Since the effective permittivity and/or permeability could be tailored to reach impossible values beyond the nature, such metamaterials have powerful abilities to manipulate electromagnetic waves and a lot of new physics has been verified, such as the negative refraction[4,5], perfect imaging[6], and invisible cloaking[7-10]. Besides the extreme medium parameters required by new physics, gradient refractive indexes have also been realized by drilling air holes or printing metallic elements on dielectric substrates, resulting in various functional devices[11-14]. However, these approaches rely on the accumulation of phase delay during wave propagation, which are always associated with large thickness and pose great challenges on fabrication at terahertz frequencies. Therefore, reducing the thickness of bulk metamaterial to that of a surface, which is called as metasurface, will be highly useful since it takes much less physical space and can be bent and twisted at will[15].

To overcome the above-mentioned limitations, the concept of generalized Snell's laws of reflection and refraction was proposed by introducing abrupt phase changes on an ultrathin metasurface[16,17]. Then a metasurface consisting of an array of C-shaped antennas has been presented to control both the phase and amplitude profiles of the transmitted waves[18], which can be utilized to design complex terahertz devices such as the dual-polarity plasmonic metalens[19] and holograms[20-22]. Other approaches such as the Y-shaped nano-antenna[23], complementary structure of V-shaped antenna[24], and graphene-loaded plasmonic antenna[25] have been suggested in metasurfaces to offer more freedom in designing the desired wavefronts. Due to the low loss, ultrathin thickness and conformal capability, metasurfaces have shown great promise in a wide range of applications from microwave[26-28], through terahertz[29,30] and infrared[31], into visible[19,32,33] frequencies.

Besides effective medium parameters, an alternative approach was recently proposed to describe metamaterials using coding sequences of '0' and '1' elements, termed as coding metamaterial[34], which can result in digital metamaterial and programmable metamaterial[35] by controlling the '0' and "1" elements digitally. Various functionalities such as steering, bending, focusing, and randomly scattering of electromagnetic waves can be realized by encoding the metasurfaces. More recently, the above idea has been extended to terahertz frequencies to manipulate terahertz waves by designing the coding particles using Minkowski fractal elements[36] and circular-ring elements[37]. In the above-mentioned coding schemes, however, the coding behaviors are independent of polarization status of electromagnetic waves due to the isotropic geometry of unit cells[35-37]. Hence they are isotropic coding metamaterials.

Here, we propose a concept of anisotropic coding metamaterial that has distinct coding behaviors for different polarizations, and demonstrate the first polarization-controlled coding metasurface, as sketched in Figs. 1a and b. In this specific illustration, under the normal (-z-direction) incidence of terahertz waves, there will be anomalous reflection to the right side when the incident field is *x*-polarized; while there will be anomalous reflection to the left side when the incident field is *y*-polarized. In fact, the polarization control through anisotropic coding is flexible and independent, and distinct functionalities can be realized for different polarizations. We present two types of unit cells, a metallic square and a dumb-bell shaped metallic structure, as the coding elements for the anisotropic coding metamaterials, which can independently reflect the normally incident terahertz waves with either 0 °(state '0') or 180 °(state '1') phases under two orthogonal polarizations. The same unit cells are then utilized to build up 2-bit anisotropic coding metasurface, in which four different digital states '00' (0 °), '01' (90 °), '10' (180 °), and '11' (270 °) are independently implemented under different polarizations. We demonstrate the polarization-controlled bifunctional performance of the encoded metasurface by far-field scattering patterns and near-electric-field distributions. Two different terahertz time-domain spectroscopy (THz-TDS) systems are employed to verify the anomalous deflections and random scattering experimentally.

## Results

### 1-bit anisotropic coding metamaterials

We start with the 1-bit case to demonstrate the control of functionalities of the anisotropic coding metasurface by polarization. An elaborately designed dumbbell-shaped metallic structure is presented as the unit cell of the anisotropic coding metasurface, as illustrated in Fig.

1c, which shows two different coding states, '1' and '0', under the normal incidences of *x*- and *y*-polarized terahertz waves. Four parameters are used to characterize the geometry of the dumbbell-shaped structure, which are heights $h_1$ and $h_2$, and widths $w_1$ and $w_2$. The structure is printed on a polyimide dielectric substrate with opaque metallic ground to provide total reflections. We show that arbitrary reflected phases can be obtained independently under the *x*- and *y*-polarizations by controlling such four parameters. Dielectric constant ε=3.0 and loss tangent δ=0.03 is set in simulations for the polyimide we use. Gold with 200 nm thickness is selected as the metallic layer. Other geometrical parameters for the 1-bit anisotropic coding element are given as: $h_1$= 45μm, $h_2$=20μm, $w_1$=37.5μm, and $w_2$=18.5μm, which are designed to operate around 1 THz. From numerical simulation results of reflection phases (Fig. 1d) by CST Microwave Studio, we note that the anisotropic unit cell (Fig. 1c) provides opposite coding states with around 180° phase difference in a wide frequency band (exactly 180° at 1 THz) under the normal incidences of terahertz waves: state '1' with the *x*-polarization and state '0' with the *y*-polarization. Hence we name the anisotropic unit cell as '1/0' coding particle. Similarly, when the anisotropic unit cell is rotated by 90° along the *z*-axis, it will produce "0" and "1" coding states for the *x*- and *y*-polarized incident terahertz waves (see Supplementary Fig. S1a), resulting in the '0/1' coding particle.

In order to control the 1-bit coding states by polarization flexibly, we still need '0/0' and '1/1' coding particles, which can be easily implemented by the isotropic unit cells[35,36]. Here, we choose the metallic square structure (Supplementary Fig. S1c)[35] with length *a* printed on the Polyimide substrate with period *p*=50 μm and thickness *d*=20 μm. When *a*=45 μm, the structure provides the '0/0' coding state for both polarizations; when *a*=30 μm, the structure provides the '1/1' coding state (Supplementary Fig. S1d). In fact, the square can be seen as a degenerated form of the dumbbell-shaped structure with $w_1=w_2$ and $h_1=w_1=a$. Hence we conclude that the reflection phases of the *x*-and *y*-polarized waves can be controlled independently by altering the four geometrical parameters ($h_1$, $h_2$, $w_1$ and $w_2$). By arranging such four coding particles ('0/0', '0/1', '1/0', and '1/1') with certain coding sequences, arbitrary functionalities can be realized in either polarization independently.

For quantitative illustration of the anisotropic coding metamaterials, we first encode a metasurface with the coding sequence of '010101...' under both *x*- and *y*-polarizations, which

is formed by repeating a two-dimensional (2D) coding matrix

$$\mathbf{M}_1^{1-\text{bit}} = \begin{pmatrix} 0/0 & 0/1 \\ 1/0 & 1/1 \end{pmatrix}.$$

We adopt a so-called "super unit cell", which is generated by a sub-array of the same basic unit cells with size $N \times N$, to minimize the unwanted coupling effect resulting from adjacent units with different geometries. Smaller super unit cell is preferred considering the minimum reflected angle allowed in our measurement system. Here, the size of super unit cell is chosen as $4 \times 4$. The deflected angle can be then calculated as $\theta = \arcsin(\lambda/\Gamma) = 48.5°$, in which $\lambda$ and $\Gamma$ represent the free-space wavelength (300 μm at 1 THz) and period of gradient unit (400 μm), respectively. A final encoded metasurface board including $16 \times 16$ super unit cells (Fig. 2a) is built up for numerical simulations. Figs. 3a and b show the three-dimensional (3D) far-field scattering patterns of the anisotropic coding metasurface under the normal incidence with $x$- and $y$-polarizations at 1 THz. We clearly see that the main lobe deviates from the $z$-axis to 48° in the $y$-$z$ plane for the $x$-polarized terahertz waves; while in the $x$-$z$ plane for the $y$-polarized terahertz waves. The deviation angle agrees very well to the analytical prediction. The bistatic scattering curves in the $y$-$z$ and $x$-$z$ planes for the $x$- and $y$-polarizations are exactly the same, as presented in Supplementary Fig. S2a. To interpret the physical insight, we also give the scattered electric-field distributions $E_x$ and $E_y$ under the $x$- and $y$-polarizations, respectively, on the $y$-$z$ and $x$-$z$ cutting planes in the near-field region, as illustrated in Supplementary Figs. S2b and c, respectively. We note that most electromagnetic energy is deflected to two symmetrically oriented directions in planes orthogonal to the polarization directions. We attribute the slight disturbance observed in the scattered electric-field to the real reflection phases with adjacent super-unit-cell coupling, which are not exactly 0° and 180° at 1 THz. However, this effect can be suppressed if we increase the size of super unit cell.

When the incident terahertz beam is polarized by 45° angle with respect to the $x$-axis, it will be equally split to four beams symmetrically oriented at the same angle with respect to the $z$-axis, as shown in Fig. 3c. This phenomenon is further verified by the near-electric-field distributions depicted in Fig. 3d, in which $E_x$ and $E_y$ are plotted in the $y$-$z$ and $x$-$z$ cutting planes, respectively. Hence, we conclude that the intensities of these four deflected beams can be arbitrarily controlled by adjusting the polarization direction of the incident electric field, promising interesting applications such as terahertz beam splitters.

The 1-bit anisotropic coding metamaterial can also be designed to deflect the *x*-polarized terahertz beam to two symmetrical directions with coding sequence 010101…, but randomly redirect the *y*-polarized beam to all directions (i.e. diffusions) with an optimized '0' and '1' coding sequence (Fig. 2b). In this case, the 3D far-field scattering patterns under the *x*- and *y*-polarizations are given in Figs. 3e and f, respectively. Although the coding pattern is different from that shown in Fig. 2a, the incident beam is still deflected as two equal beams pointing to angles $\theta=\pm48°$ with respect to the *z*-axis in the *y-z* plane under the *x*-polarization (Fig. 3g). However, for the *y*-polarization, numerous small lobes pointing to different directions are observed in Fig. 3f, indicating that the terahertz scattering of a metallic plate can be significantly reduced by covering such encoded metasurfaces. We remark that the diffused scattering waves are caused by the destructive inference resulting from the random phase difference of super unit cells. To characterize the scattering reduction performance, we give the scattering gain (in dB scale) in Fig. 3h, which is defined as the back scattering from the encoded metasurface normalized by that of the bare metallic board with the same dimension. In spite of the original single frequency design at 1 THz, it is able to reduce the scattering level significantly by at least -10 dB in the frequency range from 0.9 to 1.5 THz. This example demonstrates the powerful abilities of the proposed anisotropic coding metasurface to manipulate the *x*- and *y*-polarized incident terahertz waves independently.

**2-bit anisotropic coding metamaterials**

Based on the design method of 1-bit anisotropic coding metamaterial, we further introduce 2-bit anisotropic coding metamaterial, which can provide more flexibility in manipulating the terahertz waves. The basic unit cells of 2-bit coding is able to reflect the incident waves with phase responses of 0°, 90°, 180° and 270°, corresponding to coding elements '00', '01', '10', and '11', respectively. Since the coding elements are designed to exhibit four distinct states independently under *x*- and *y*-polarizations, a total number of 16 basic unit cells are required for the 2-bit anisotropic coding metamaterial. The structures of these 16 unit cells are shown in Fig. 1E, whose geometrical parameters are given in Supplementary Table S1. In the 2-bit case, the thickness of the polyimide layer is chosen as 25 μm. Here, the metallic square, rectangular, and dumbbell-shaped structures are used to build up the four isotropic and 12 anisotropic coding elements. The six anisotropic elements in the upper triangular area can be obtained by rotating those in the lower triangular area by 90° in the *x-y* plane (see Fig. 1e).

To demonstrate the flexibility of 2-bit anisotropic coding metamaterials in manipulating terahertz waves, we first encode a metasurface with the same coding sequence '00-01-10-11-00-01-10-11…' for both polarizations, in which the size of the super unit cell is 2×2. In this case, the 2D coding matrix is written as

$$\mathbf{M}_1^{2-bit} = \begin{pmatrix} 00/00 & 00/01 & 00/10 & 00/11 \\ 01/00 & 01/01 & 01/10 & 01/11 \\ 10/00 & 10/01 & 10/10 & 10/11 \\ 11/00 & 11/01 & 11/10 & 11/11 \end{pmatrix},$$

which generates the final 2-bit coding layout shown in Fig. 2c. Since each super unit cell has gradient phase difference of 90°, based on the generalized Snell's law[17], the normally incident beam will be reflected to the angle 48° in the plane orthogonal to the polarization direction (Supplementary Fig. S2d). This phenomenon can be clearly observed from the 3D far-field scattering patterns presented in Figs. 4a and b. For the *x*-polarized normal incidence, the terahertz beam is anomalously reflected to the direction of ($\varphi$=90°, $\theta$=48°) in the *y-z* plane; for the *y*-polarization, however, the terahertz beam is directed to the angle ($\varphi$=180°, $\theta$=48°) in the *x-z* plane. The flexible control of terahertz reflections by polarization is further verified by the near-field distributions shown in Figs. 4c and d. Similar to the 1-bit case, the inhomogeneity modifies the mutual coupling between neighboring unit cells and response to the incident wave, thus resulting in the observable disturbance of scattered fields.

Particularly, the 2-bit anisotropic coding metamaterial is utilized to design a terahertz free-background reflected quarter-wave plate, which can produce a circularly polarized beam that bends away from the surface normal when a normally incident wave is linearly polarized by 45° with respect to the *x*-axis. By designing the reflection phase difference of each unit cell under the *x*- and *y*-polarizations to be 90°, a normal reflection of equal reflection amplitude with 90° phase difference is obtained when the electric field is along the 45° polarization, resulting in a circularly polarized wave. If we arrange these unit cells by a phase gradient along one direction, the circularly polarized wave will be reflected to an oblique angle, thus forming a background-free circularly polarized wave. Such functionality is realized by encoding the metasurface with the coding matrix

$$\mathbf{M}_2^{2-bit} = \begin{pmatrix} 00/01 & 01/10 & 10/11 & 11/00 \\ 00/01 & 01/10 & 10/11 & 11/00 \\ 00/01 & 01/10 & 10/11 & 11/00 \\ 00/01 & 01/10 & 10/11 & 11/00 \end{pmatrix}.$$

In this case, the size of super unit cell is selected as 3×3, and the corresponding 2-bit coding metasurface layout is illustrated in Fig. 2d. The 3D far-field scattering pattern shown in Fig. 4e reveals that the normally incident linearly-polarized terahertz beam is anomalously reflected to 30 ° (see Supplementary Fig. S3a for details) in the *x-z* plane (φ=180°, θ=30°) with axial ratio of only 1.03 at 1THz (see Fig. 4f), indicating an almost ideal circularly-polarized wave. Away from the designed frequency, the phase and amplitude responses of each unit cell will deviate from their original values. However, the well-designed reflected quarter-wave plate still works in a broad frequency band, as shown by the blue curve in Fig. 4f. The reflected angle linearly decreases from 38 ° to 24.5 ° as frequency goes from 0.8 to 1.2 THz, while the axial ratio remains below 1.26 in the entire bandwidth. Supplementary Fig.S3b further gives the axial ratio from -45 ° to -15 ° in the *x-z* plane, from which we observe axial ratios lower than 1.15 from -35 ° to -25 °. The reflected quarter-wave plate with outstanding performance can be used as high-efficiency circular polarizer and may find interesting applications in terahertz systems.

More manipulations of terahertz waves by 2-bit anisotropic coding metamaterials under different polarizations are given in Supplementary Information and Figs. S4 and S5.

**Experiments and measured results**

To experimentally validate the performance of the anisotropic coding metasurfaces, three samples (two 1-bit and one 2-bit) are fabricated using standard photolithography processes, as demonstrated in Figs. 5a and b, and Supplementary Figs.S6a and b, respectively, which correspond to the coding layouts shown in Figs. 2b, a, and c. Each sample covers area of 15×15 mm$^2$ to accommodate the plane-wave like incident beam, whose diameter is measured around 3 mm. The freestanding sample (Fig. 5a) is so flexible that can be wrapped on any objects with curved surfaces, predicting promising applications such as conformal cloaking[38] at terahertz frequencies. Fig. 5b illustrates the zoomed image of the 1-bit coding sample (Fig. 2b) taken by an optical microscope (VHX-5000, Keyence Company). Two terahertz measurement systems, rotational THz-TDS and theta-to-theta THz-TDS, were employed to measure the deflected angle and diffused scattering of the fabricated sample, respectively.

The measured reflection amplitudes with respect to receiving angle of the 1-bit coding metasurface encoded with coding matrix $\mathbf{M}_1^{1-\text{bit}}$ (Supplementary Fig. S6a) under *x* and *y* polarizations are given in Fig. 6a, which are in the *y-z* and *x-z* planes, respectively. Note that all measurement data were plotted at 1 THz after data smoothing treatment to eliminate the background noise. We clearly see two nearly identical peaks between 40° to 60° with the maximum values appearing at around 51° for both polarizations, indicating that the normally incident terahertz beam is reflected to the angle 51° under both polarizations. We remark that since the angles of two deflected beams in this design are symmetrical with respect to the *x-z* or *y-z* plane, only the deflected angles from 0° to 90° are measured in experiments. The measured result (51°) has very good agreement with theoretical result (48.5°). We note that the maximum amplitudes of anomalous reflections (between 0.5 and 0.53) from the above measurements are slightly larger than simulations (0.456). This discrepancy could be resulted from the inaccurate thickness of the PI layer and the inevitable tolerance in the dimensions of structure during fabrication, both of which could lead to distorted scattering pattern that may result in higher or lower radiations in certain directions.

For the other 1-bit anisotropic coding metasurface encoded with coding matrix $\mathbf{M}_2^{1-\text{bit}}$ (Fig. 5b), we give the measured reflection amplitude versus receiving angle in Fig. 6b when it is normally illuminated with the *x*-polarized terahertz beam. A similar scattering pattern to Fig. 6a is obtained despite the fact that the coding sequences are totally different. When the polarization of incident beam turns 90° to the *y*-axis, however, the normally incident beam is evenly scattered to a lot of directions in the upper half space, resulting in low reflection in specular reflection directions, which can be verified by the scattering gains under different specular reflection angles at 0°, 20°, 40° and 60° in Fig. 6c. The reflections under normal incidence remain below -10 dB from 0.9 to 1.2 THz, and this low-reflection band is further expanded to 0.5 THz (from 0.88 to 1.38 THz) under the specular reflection angle of 20°. The random scattering performance deteriorates for larger specular reflection angle, because the current measurement setup only supports transverse-magnetic polarization (the electric field is horizontally polarized). At oblique incidences, the wave vector is no longer parallel to the unit-cell plane, which will lead to the appearance of higher-order modes and influence the reflection responses of unit cells [39].

For the 2-bit anisotropic metasurface encoded by matrix $\mathbf{M}_1^{2-\mathrm{bit}}$ (Supplementary Fig. S6b), the measured reflection amplitudes from 25° to 90° for the *x*-polarization (in the *y-z* plane) and *y* polarization (in the *x-z* plane) are illustrated in Fig. 6d, in which we can clearly observe that the normally incident beam is reflected to 52° angle in the *y-z* plane (for the *x*-polarization) and in the *x-z* plane (for the *y*-polarization). Fig. 6e further gives the measured reflection amplitudes from 0.4 to 1.8 THz of the 2-bit case (same pattern with Fig. 6d) under the *x*-polarization when the receiver scans from 25° to 90°. Significant anomalous reflection peaks are detected from 35° to 75°. We note that these reflection peaks shift to smaller angles with the increase of frequency, as expected by the general Snell's law. The good agreements between experimental and simulation results clearly demonstrate both the functionalities and performances of the 1-bit and 2-bit anisotropic coding metamaterials.

## Discussions

1-bit and 2-bit anisotropic coding metamaterials composed of squares and dumb-bell-shaped structures have been demonstrated theoretically and experimentally at terahertz frequencies to exhibit multiple functionalities under different polarizations. We showed powerful capability of the anisotropic coding metamaterials in controlling terahertz waves by coding sequence and polarization independently. For examples, we illustrated that a 1-bit anisotropic coding metasurface produces anomalous reflections along directions $\theta=\pm\arcsin(\lambda/\Gamma)$ from the surface normal under the incidence of *x*-polarized terahertz waves; under the *y*-polarization, however, the same anisotropic coding metasurface forms diffused scattering waves in the upper half space. Owing to the ultrathin and flexible nature of the proposed design, this technique could be employed to reduce the monostatic and bistatic scattering of a curved object effectively by simply warping the metasurface on it. In 2-bit anisotropic coding metasurface, by designing a 90° phase difference under the *x*- and *y*-polarizations for each unit and arranging the units with gradient phase along a pre-designed direction, we demonstrated a free-background reflected quarter-wave plate that is capable of converting normally incident linearly-polarized terahertz waves to obliquely reflected circularly-polarized waves with high efficiency. We believe that the outstanding functional performance and good features of design (e.g. ultrathin thickness, flexibility, and easy fabrication) will make the proposed anisotropic coding metasurface more attractive for low-cost terahertz devices and circuits.

## Methods

### Sample fabrication

In fabricating the terahertz anisotropic coding metasurface, a 180-nm-thick gold layer was deposited on a silicon wafer by electron beam evaporation to serve as the metallic background. Then, spin coating process enables the uniform deposition of the liquid polyimide (Yi Dun New Materials Co. Ltd, Suzhou) on the gold layer, which was then solidified on a hot plate at 80, 120, 180 and 250 °C for 5, 5, 5 and 20 minutes, respectively. Considering the viscosity of polyimide and the minimum spin speed, these spin-coating and curing processes need to be repeated for two and three times to complete the final polyimide layers with 20 and 25 μm thicknesses, respectively. Next, followed by the standard photolithography, another Ti/Au layer (10/180 nm) was deposited by electron beam evaporation, and then a lift-off process was used to form the final metallic pattern. The sample can be easily peeled off from the wafer substrate owing to the poor adhesion between the gold and silica layers (grown on the silicon wafer).

### Measurement systems

A sketch of the experimental setup of rotational THz-TDS system is presented in Fig. 5c. Here, a pair of fiber-based terahertz photoconductive switches (Model TR4100-RX1, API Advanced Photonix, Inc.) was used to generate and detect the time- domain terahertz pulses. The position of transmitter was fixed while the receiver was placed on a strip-shaped holder, which can be rotated around the central metallic cylinder (see Supplementary Fig. S6c). The sample was attached to a metallic board mounted on the central cylinder. The distances between the transmitter and receiver to the sample were kept the same as 23 cm and the angle between them could be scanned from 24° to 90° in the *y-z* plane. Both the transmitter and receiver were optically excited by a commercial ultrafast erbium fiber laser system (T-Gauge, API Advanced Photonix, Inc.), whose available spectrum ranges from 0.3 to 3.0 THz. The deflected beam scattered from the sample was recorded every 5 ° from 25 ° to 90 °, except at 80 ° and 85 ° because the signals measured at such large oblique angles are very weak. The signal reflected from an opaque gold film was first recorded as a reference for all incident angles.

The schematic diagram of the experimental setup for the theta-to-theta THz-TDS system (Zomega Z-3) is presented in Fig. 5d. The sample, attached to a sample holder, was located on a rotational stage, and hence could be automatically rotated from 20 ° to 90 °. A parabolic mirror located on another independent rotational stage was used to normally reflect the terahertz beam (reflected from the sample) back to sample because the parabolic mirror could automatically aim at the specular reflection direction of the incident beam. The signal was then reflected to the receiver by several mirrors. In this measurement, we recorded the terahertz wave reflected from the opaque gold film at 0 °, 20 °, 40 °, and 60 ° as references. Detailed description of the theta-to-theta THz-TDS system can be found in Supplementary Fig. S6d.

## References

1   D. R. Smith, W. J. Padilla, D. C. Vier, S. C. Nemat-Nasser, S. Schultz, Composite medium with simultaneously negative permeability and permittivity. *Phys. Rev. Lett.* **84**, 4184-4187 (2000).

2   R. P. Liu, T. J. Cui, D. Huang, B. Zhao, D. R. Smith, Description and explanation of electromagnetic behaviors in artificial metamaterials based on effective medium theory. *Phys. Rev. E* **76**, 026606 (2007).

3   D. R. Smith, S. Schultz, Determination of effective permittivity and permeability of metamaterials from reflection and transmission coefficients. *Phys. Rev. B* **65**, 195104 (1999).

4   R. A. Shelby, D. R. Smith, S. Schultz, Experimental verification of a negative index of refraction. *Science* **292**, 77-79 (2001).

5   H. Shin, S. H. Fan, All-angle negative refraction for surface plasmon waves using a metal-dielectric-metal structure. *Phys. Rev. Lett.* **96**, 073907 (2006).

6   J. B. Pendry, Negative refraction makes a perfect lens. *Phys. Rev. Lett.* **85**, 3966-3969 (2000).

7   R. Liu, et al. Broadband Ground-Plane Cloak. *Science* **323**, 366-369 (2009).

8   D. Schurig, et al. Metamaterial electromagnetic cloak at microwave frequencies. *Science* **314**, 977-980 (2006).

9   F. Magnus, et al. A D.C. magnetic metamaterial. *Nat Mater.* **7**, 295-297 (2008).

10  H. F. Ma, T. J. Cui, Three-dimensional broadband ground-plane cloak made of metamaterials. *Nat Commun.* **1**, 21 (2010).

11  T. J. Cui, D. R. Smith, R. Liu, Metamaterials: Theory, Design, and Applications (Springer, 2009).

12  H. F. Ma, T. J. Cui, Three-dimensional broadband and broad-angle transformation-optics lens. *Nat Commun.* **1**, 124 (2010).

13  A. V. Kabashin, et al. Plasmonic nanorod metamaterials for biosensing. *Nat Mater.* **8**, 867-871 (2009).

14  Q. Cheng, T. J. Cui, W. X. Jiang, B. G. Cai, An omnidirectional electromagnetic absorber made of metamaterials. *New J Phys* **12**, 063006 (2010).

15  C. L. Holloway, et al. An Overview of the Theory and Applications of Metasurfaces: The Two-Dimensional Equivalents of Metamaterials. *IEEE Trans. Antennas Propag.* **54**, 10-35 (2012).

16  F. Aieta, et al. Out-of-Plane Reflection and Refraction of Light by Anisotropic Optical Antenna Metasurfaces with Phase Discontinuities. *Nano Lett.* **12**, 1702-1706 (2012).

17  N. F. Yu, et al. Light Propagation with Phase Discontinuities: Generalized Laws of Reflection and Refraction. *Science* **334**, 333-337 (2011).

18  L. X. Liu, et al. Broadband Metasurfaces with Simultaneous Control of Phase and Amplitude. *Adv Mater.* **26**, 5031-5036 (2014).

19  X. Z. Chen, et al. Dual-polarity plasmonic metalens for visible light. *Nat Commun.* **3** 1198 (2012).

20  G. X. Zheng, H. M. Mitchell Kenney, G. X. Li, T. Zentgraf, S. Zhang, Metasurface holograms reaching 80% efficiency. *Nat Nanotechnol.* **10**, 308-312 (2015).

21  W. T. Chen, et al. High-Efficiency Broadband Meta-Hologram with Polarization-Controlled Dual Images. *Nano Lett.* **14**, 225-230 (2014).

22  X. J. Ni, A. V. Kildishev, V. M. Shalaev, Metasurface holograms for visible light. *Nat Commun.* **4**, 2807 (2013).

23  M. A. Kats, et al. Giant birefringence in optical antenna arrays with widely tailorable optical anisotropy. *P Natl Acad Sci USA* **109**, 12364-12368 (2012).

24  X. J. Ni, S. Ishii, A. V. Kildishev, V. M. Shalaev, Ultra-thin, planar, Babinet-inverted plasmonic metalenses. *Light: Science & Applications* **2**, e72 (2013).

25  Y. Yao, et al. Broad Electrical Tuning of Graphene-Loaded Plasmonic Antennas. *Nano Lett.* **13**, 1257-1264, (2013).

26  Y. B. Li, X. Wan, B. G. Cai, Q. Cheng, T. J. Cui, Frequency-Controls of Electromagnetic Multi-Beam

Scanning by Metasurfaces. *Sci. Rep.* **4**, 6921 (2014).

27  J. Zhang, Z. L. Mei, W. R. Zhang, F. Yang, T. J. Cui, An ultrathin directional carpet cloak based on generalized Snell's law. *Appl. Phys. Lett.* **103**, 151115 (2013).

28  S. L. Sun, et al. Gradient-index meta-surfaces as a bridge linking propagating waves and surface waves. *Nat Mater.* **11**, 426-431 (2012).

29  X. P. Shen, et al. Triple-band terahertz metamaterial absorber: Design, experiment, and physical interpretation. *Appl. Phys. Lett.* **101**, 154102 (2012).

30  H. T. Chen, et al. A metamaterial solid-state terahertz phase modulator. *Nat Photonics.* **3**, 148-151 (2009).

31  N. F. Yu, et al. A Broadband, Background-Free Quarter-Wave Plate Based on Plasmonic Metasurfaces. *Nano Lett.* **12**, 6328-6333 (2012).

32  F. Monticone, N. M. Estakhri, A. Alu, Full Control of Nanoscale Optical Transmission with a Composite Metascreen. *Phys. Rev. Lett.* **110**, 203903 (2013).

33  X. J. Ni, N. K. Emani, A. V. Kildishev, A. Boltasseva, V. M. Shalaev, Broadband Light Bending with Plasmonic Nanoantennas. *Science* **335**, 427-427 (2012).

34  C. D. Giovampaola, N. Engheta, Digital Metamaterials. *Nat Mater.* **13**, 1115-1121 (2014).

35  T. J. Cui, M. Q. Qi, X. Wan, J. Zhao, Q. Cheng, Coding metamaterials, digital metamaterials and programmable metamaterials. *Light: Science & Applications* **3**, e218 (2014).

36  L. H. Gao, et al. Broadband diffusions of terahertz waves by multi-bit coding metasurfaces. *Light: Science & Applications* **4**, e324 (2015).

37  L. Liang, *et* al., "Anomalous terahertz reflection and scattering by flexible and conformal coding metamaterial," *Advanced Optical Materials*, DOI: 10.1002/adom.201500206 (2015).

38  S. Liu, H. X. Xu, H. C. Zhang, T. J. Cui, Tunable ultrathin mantle cloak via varactor-diode-loaded metasurface. *Opt. Express* **22**, 13403-13417 (2014).

39  Y. Q. Ye, Y. Jin, S. L. He, Omnidirectional, polarization-insensitive and broadband thin absorber in the terahertz regime. *J Opt Soc Am B* **27**, 498-504 (2010).

**Acknowledgements:** S. L. and T. J. C. contributed equally to this work. This work was supported by the National Science Foundation of China (61571117, 61171024, 61171026, 61138001, 61302018 and 61401089), and the 111 Project (111-2-05).

**Author contributions:** S.L. carried out the analytical modelling, numerical simulations, sample fabrication, and measurements. T.J.C., as the principal investigator of the projects, conceived the idea, suggested the designs, planned, coordinated and supervised the work. X.W., W.X.T., X.Y.Z., and H.Y. made part of numerical simulations, and Q.X., D.B., L.D. C.O., H.F.M., W.X.J., J.H., W.L.Z., and Q.C. took part in the sample fabrications and measurements. All authors discussed the theoretical and numerical aspects and interpreted the results. All authors contributed to the preparation and writing of the manuscript.

**Competing financial interests:** The authors declare no competing financial interests.

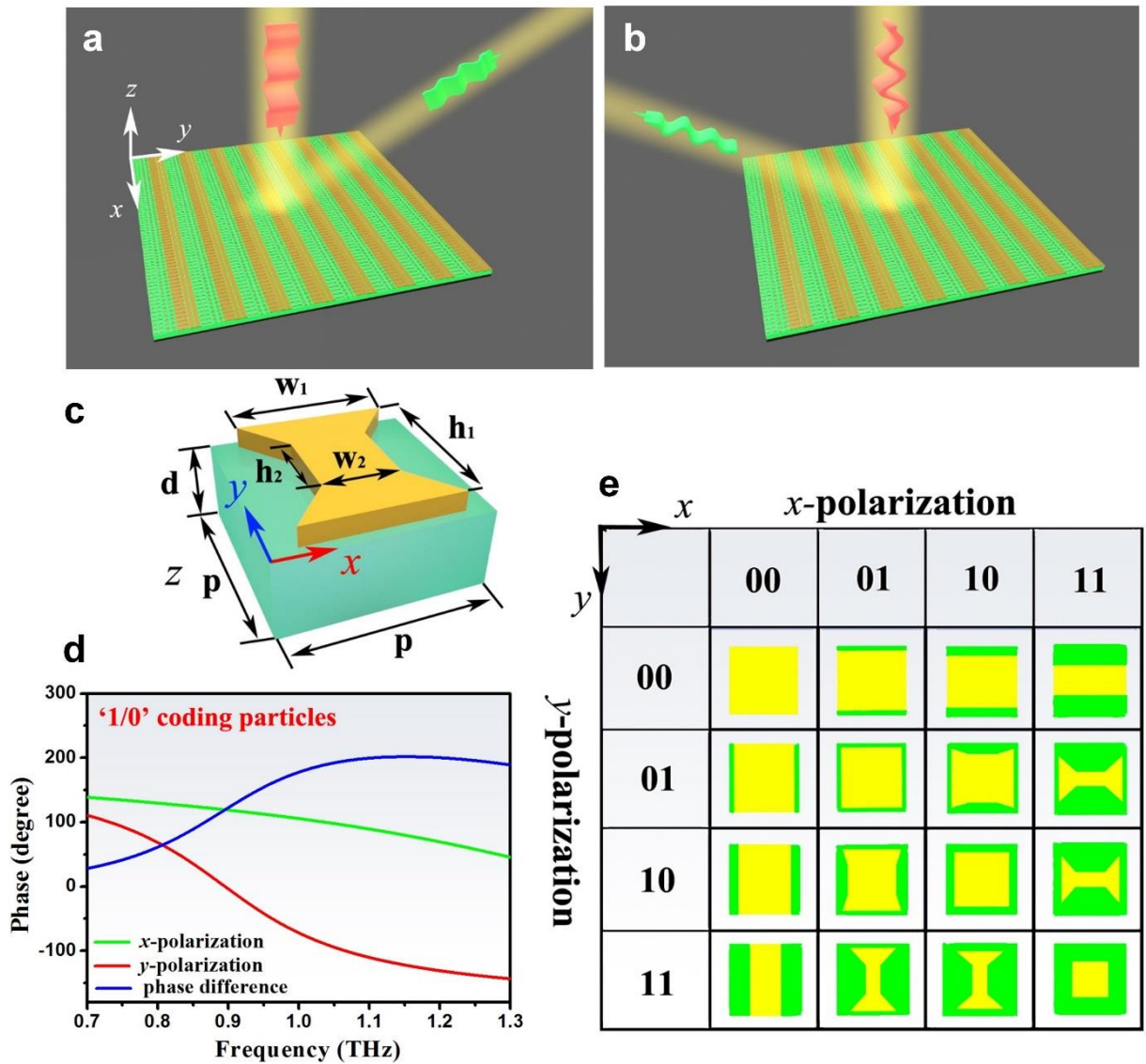

**Figure 1 | Illustration of anisotropic coding metamaterial and the structure design**. (**a,b**) An example to demonstrate the flexible ability of the encoded metasurface to anomalously reflect the normal incident beam to the right side under the *x*-polarization and the left side under the *y*-polarization. (**c**) The structure of anisotropic coding element '1/0' (without the metallic background). (**d**) Reflection phases and the corresponding phase difference for the anisotropic unit cell '1/0' under the *x*- and *y*- polarizations. (**e**) The structure of 16 unit cells for the 2-bit anisotropic coding metasurface.

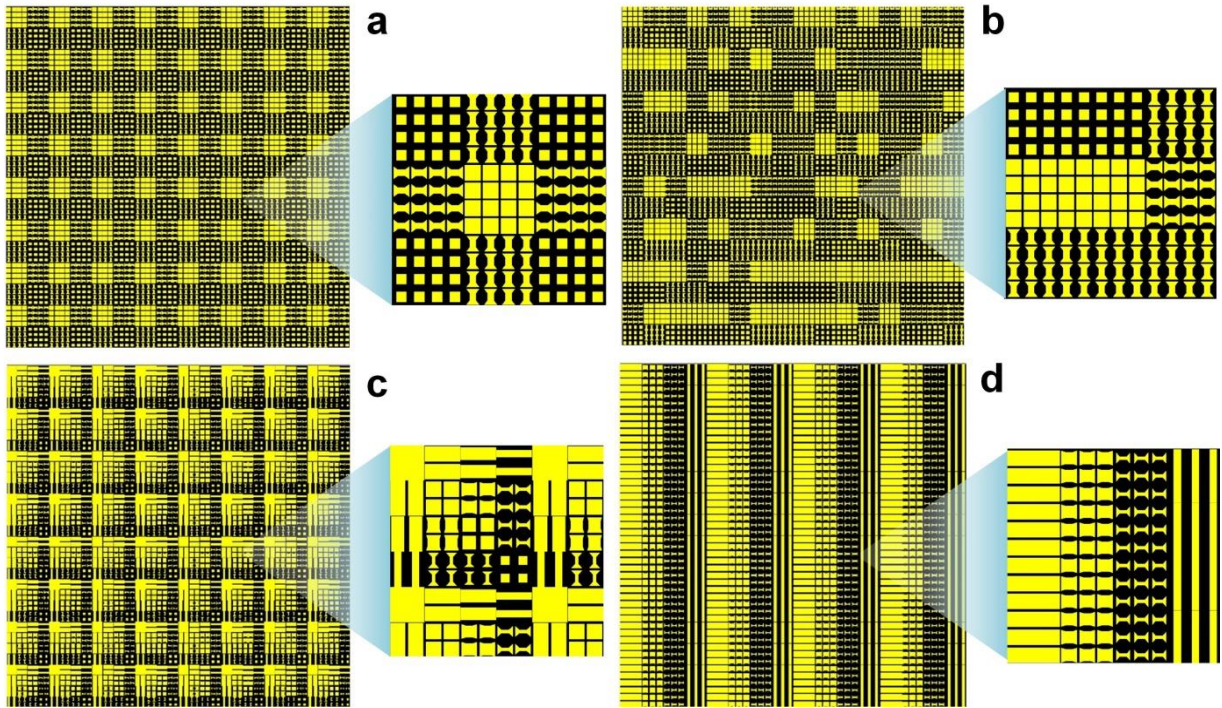

**Figure 2 | The coding patterns of 1-bit and 2-bit coding metasurfaces and their zoomed views.** (**a**) The pattern with the coding matrix $\mathbf{M}_1^{1-bit}$, which contains 16×16 super unit cells with size 4×4. (**b**) The pattern with the coding matrix $\mathbf{M}_2^{1-bit}$, which contains 16×16 super unit cells with size 4×4. (**c**) The pattern with the coding matrix $\mathbf{M}_1^{2-bit}$, which contains 32×32 super unit cells with size 2×2. (**d**) The pattern with the coding matrix $\mathbf{M}_2^{2-bit}$, which contains 16×16 super unit cells with size 3×3.

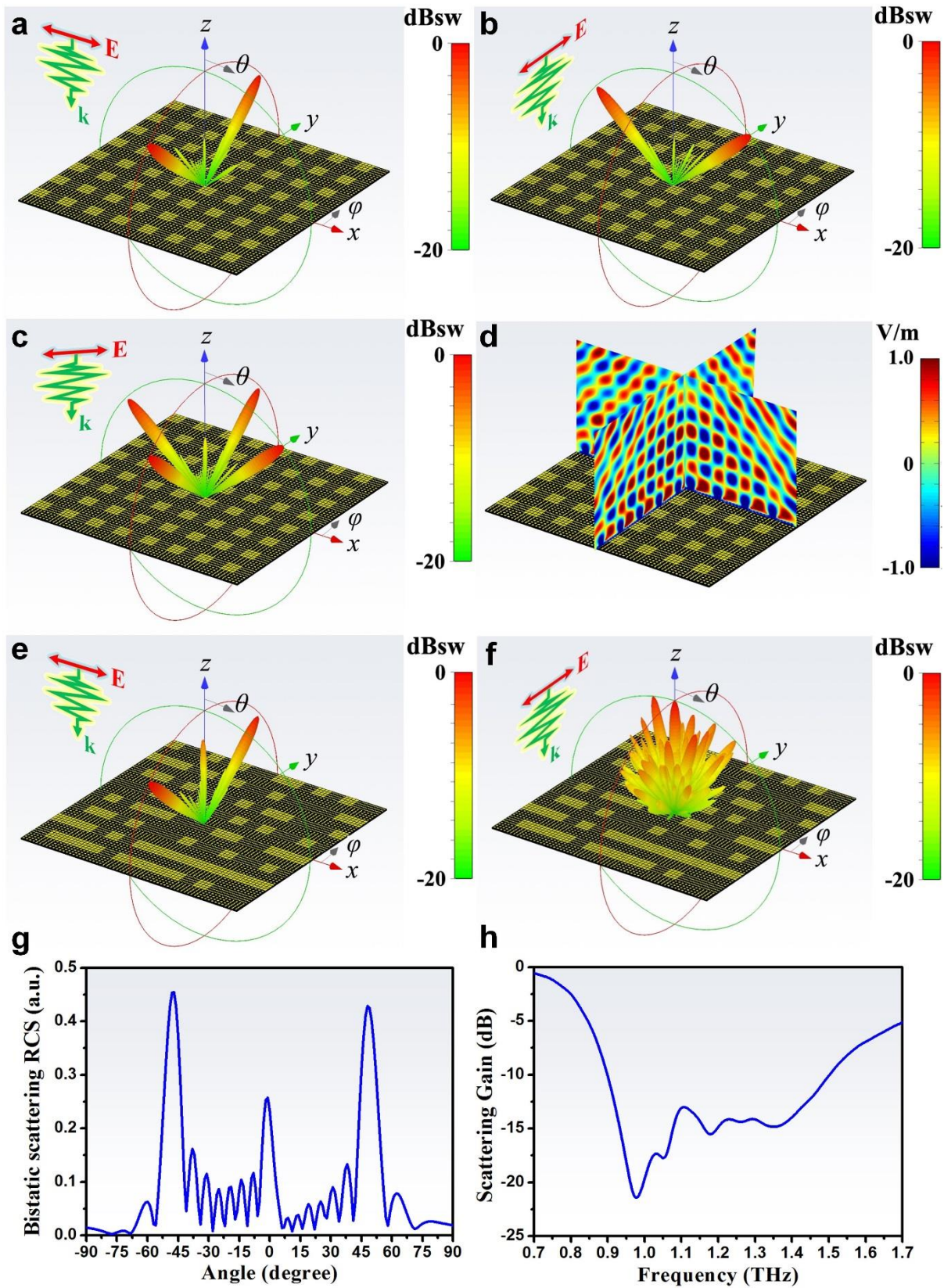

**Figure 3 | The simulated 3D and 2D scattering patterns for the 1-bit anisotropic coding metasurface. (a-c)** The 3D far-field scattering patterns of the metasurface encoded with the coding matrix $\mathbf{M}_1^{1-\text{bit}}$ under the *x*-, *y*- and 45°- (with respect to the *x*-axis) polarizations,

respectively. **(d)** The near-electric-field distributions for the same metasurface in **(a)**. The electric fields on the *y-z* and *x-z* cutting planes are plotted with the $E_x$ and $E_y$ components, respectively. **(e,f)** The 3D far-field scattering patterns of the metasurface encoded with the coding matrix $\mathbf{M}_2^{1-\text{bit}}$ under the *x*- and *y*-polarizations, respectively. **(g)** The 2D far-field scattering pattern plotted in the *y-z* plane for the same metasurface in **(e)** under the *x*-polarized illumination. **(h)** The scattering gain, defined by the reduction of back scattering compared with the bare metallic case, for the same metasurface in (**e**) under the *y*-polarization.

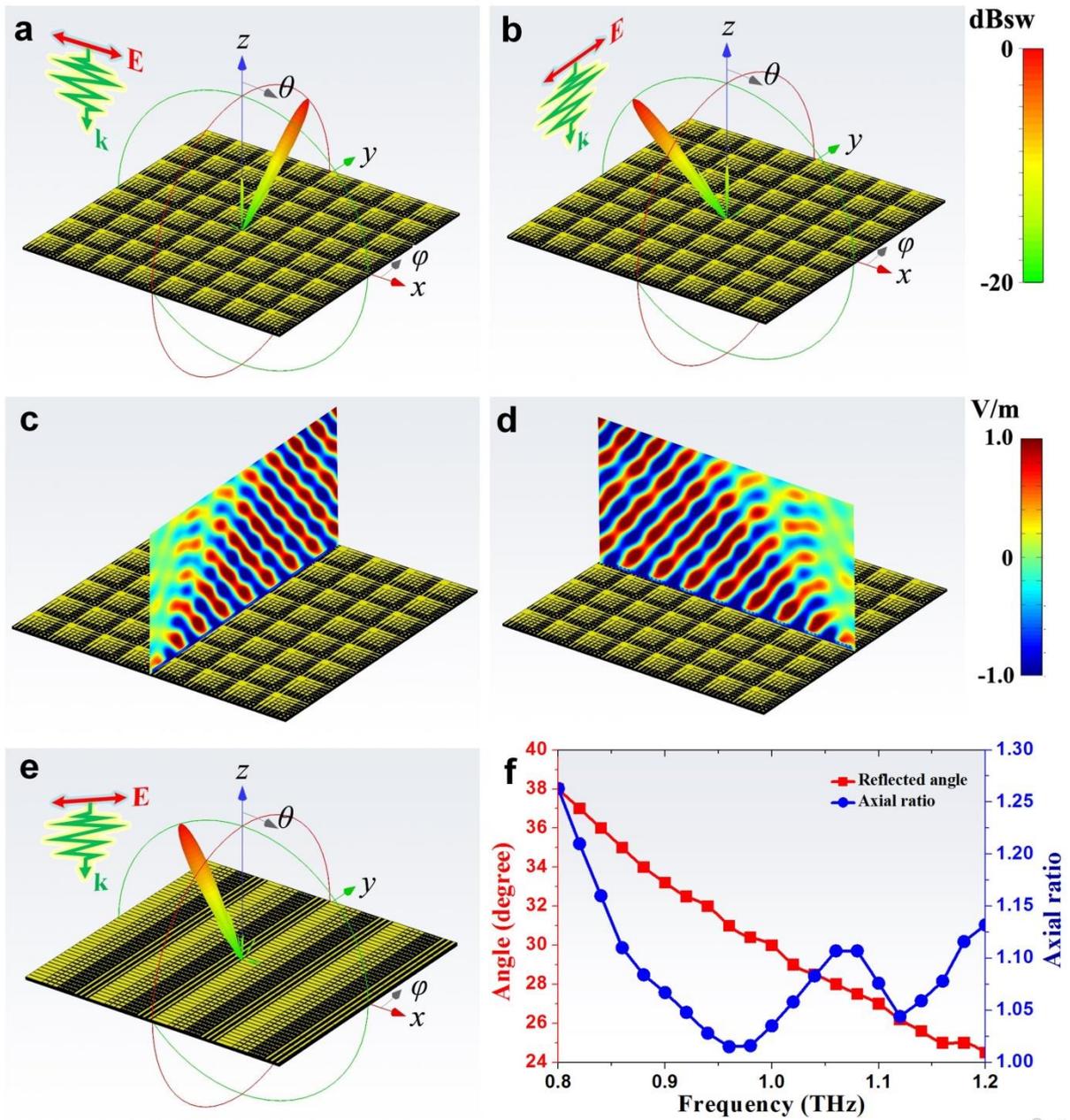

**Figure 4 | The simulated results for the 2-bit anisotropic coding metasurface. (a,b)** The 3D far-field scattering patterns for the metasurface encoded with the coding matrix $\mathbf{M}_1^{2-\mathrm{bit}}$ under the *x*- and *y*-polarizations, respectively. **(c,d)** The corresponding near-electric-field distributions $E_x$ and $E_y$ on the *y-z* and *x-z* cutting planes. **(e)** The 3D far-field scattering pattern for the metasurface encoded with the coding matrix $\mathbf{M}_2^{2-\mathrm{bit}}$ when the incident terahertz wave is linearly polarized by 45° with respect to the *x*-axis. **(f)** The variation of anomalously reflected angle (red square line) and the axial ratio (blue circle line) as the frequency ranges from 0.8 to 1.2 THz for the same metasurace in **(e)**. The axial ratio at each frequency point is obtained in the maximum scattering direction.

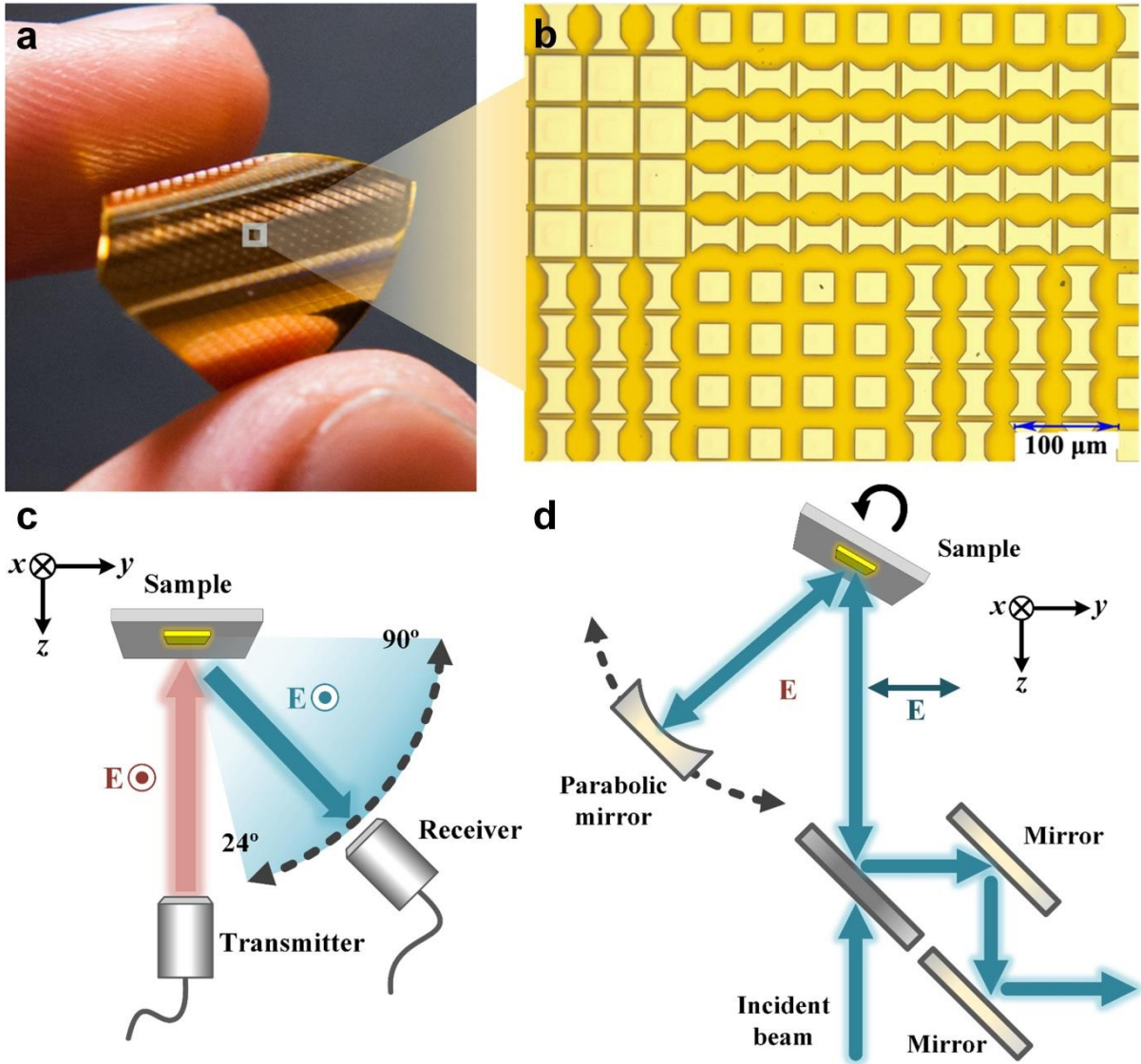

**Figure 5 | The photographs of fabricated sample and schematics of experimental setup.**
**(a)** The freestanding sample released from silicon wafer. **(b)** The optical microscopy image of the sample encoded with the coding matrix $\mathbf{M}_2^{1-\text{bit}}$. **(c,d)** The schematics of the experimental configurations for the rotational THz-TDS and theta-to-theta THz-TDS systems.

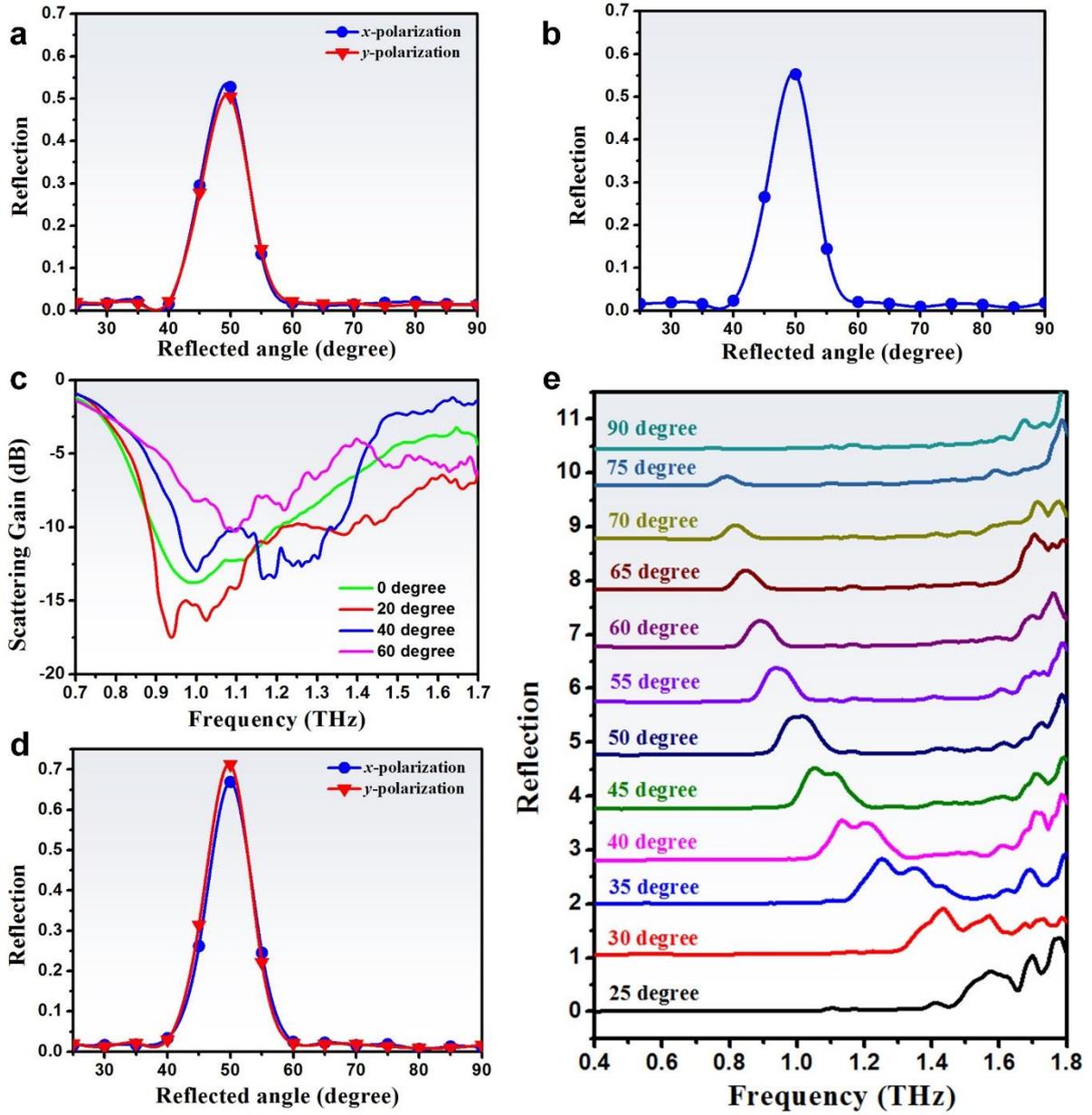

**Figure 6 | The experimental results for anisotropic coding metasurfaces encoded with three different coding matrices. (a,d)** The amplitudes of reflections versus receiving angle for the metasurfaces encoded with the coding matrices $\mathbf{M}_1^{1-\text{bit}}$ and $\mathbf{M}_1^{2-\text{bit}}$ under both polarizations. **(b)** The amplitudes of reflections with respect to receiving angle for the 1-bit anisotropic metasurface encoded with the coding matrix $\mathbf{M}_2^{1-\text{bit}}$ under the *x*-polarization. **(c)** The scattering gains for the same metasurface in **(b)** under the *y*-polarization measured at 0°, 20°, 40° and 60°. **(e)** The reflection amplitudes from 0.4 to 1.8 THz for the same metasurface in **(d)** under the *x*-polarization when the receiver scans from 25° to 90°.

Supplementary Information for

# Polarization-controlled anisotropic coding metamaterials at terahertz frequencies

Shuo Liu, Tie Jun Cui*, Quan Xu, Di Bao, Liangliang Du, Xiang Wan, Wen Xuan Tang, Chunmei Ouyang, Xiao Yang Zhou, Hao Yuan, Hui Feng Ma, Wei Xiang Jiang, Jiaguang Han, Weili Zhang, and Qiang Cheng

[1] State Key Laboratory of Millimeter Waves, Southeast University, Nanjing 210096, China
[2] Synergetic Innovation Center of Wireless Communication Technology, Southeast University, Nanjing 210096, China
[3] Cooperative Innovation Centre of Terahertz Science, No.4, Section 2, North Jianshe Road, Chengdu 610054, China
[4] Center for Terahertz Waves and College of Precision Instrument and Optoelectronics Engineering, Tianjin University, Tianjin 300072, China
[5] Jiangsu Xuantu Technology Co., Ltd., 12 Mozhou East Road, Nanjing 211111, China

* These authors contribute equally to this work.

+ Corresponding author: E-mail: tjcui@seu.edu.cn.

**This PDF file includes:**

The detailed descriptions of 1) structures and reflection responses of the basic coding particles; 2) conversion efficiency of the anisotropic coding metamaterial; 3) reflected quarter-wave plate and in-plane beam scanner; 4) experiments and measurements; 5) Supplementary Figs. S1-S6; and 6) Supplementary Table S1.

## Structures and reflection responses of the basic coding particles

Supplementary Fig. S1a gives the structure of the '0/1' coding particle, which is obtained by rotating the '1/0' coding particle by 90° in the *x-y* plane. This can be verified by the reflection responses under the normally incident *x*- and *y*-polarized terahertz waves shown in Supplementary Fig. S1b, in which the curves under both polarizations are identical to Main text Fig. 1d except that they are interchanged with each other. The structure of the isotropic unit cell is presented in Supplementary Fig. S1c, whose reflection responses for both '0/0' and '1/1' coding particles are given in Supplementary Fig. S1d. The geometrical parameters for both plots are the same as those given in the main text for the 1-bit case. By observing the phase difference in both cases, we find that it varies from 160° to 180° (20° tolerance) for the isotropic coding particle from 0.86 to 1.17 THz, covering a relative bandwidth of 30.5%. For the isotropic coding particle, the relative bandwidth with 21° phase difference tolerance (from 160° to 201°) increases to 32.1% (from 0.94 to 1.7 THz). The broadband design of the basic coding elements gives rise to excellent performance of encoded metasurfaces such as broadband diffusions (Main text Fig. 3h) and wide-angle scanning (Main text Figs. 4f and 6e).

**Supplementary Table S1 | Geometrical parameters of unit cells for the 2-bit anisotropic coding metamaterials.** For clarity, each parameter is marked by the geometrical parameter (the first item) and the coding state (the second item).

| Parameters | A_00 | $W_1$_10 | $W_2$_10 | $H_1$_10 | $H_2$_10 |
|---|---|---|---|---|---|
| Value (μm) | 50 | 43.5 | 43.5 | 50 | 20 |
| Parameters | A_11 | $W_1$_20 | $W_2$_20 | $H_1$_20 | $H_2$_20 |
| Value (μm) | 44 | 37 | 37 | 50 | 20 |
| Parameters | A_22 | $W_1$_30 | $W_2$_30 | $H_1$_30 | $H_2$_30 |
| Value (μm) | 38.5 | 21 | 21 | 50 | 20 |
| Parameters | A_33 | $W_1$_21 | $W_2$_21 | $H_1$_21 | $H_2$_21 |
| Value (μm) | 25 | 42 | 34.5 | 44 | 20 |
| Parameters |  | $W_1$_31 | $W_2$_31 | $H_1$_31 | $H_2$_31 |
| Value (μm) |  | 32 | 10 | 44 | 20 |

| Parameters | W₁_32 | W₂_32 | H₁_32 | H₂_32 |
|---|---|---|---|---|
| Value (μm) | 32 | 10 | 39 | 20 |

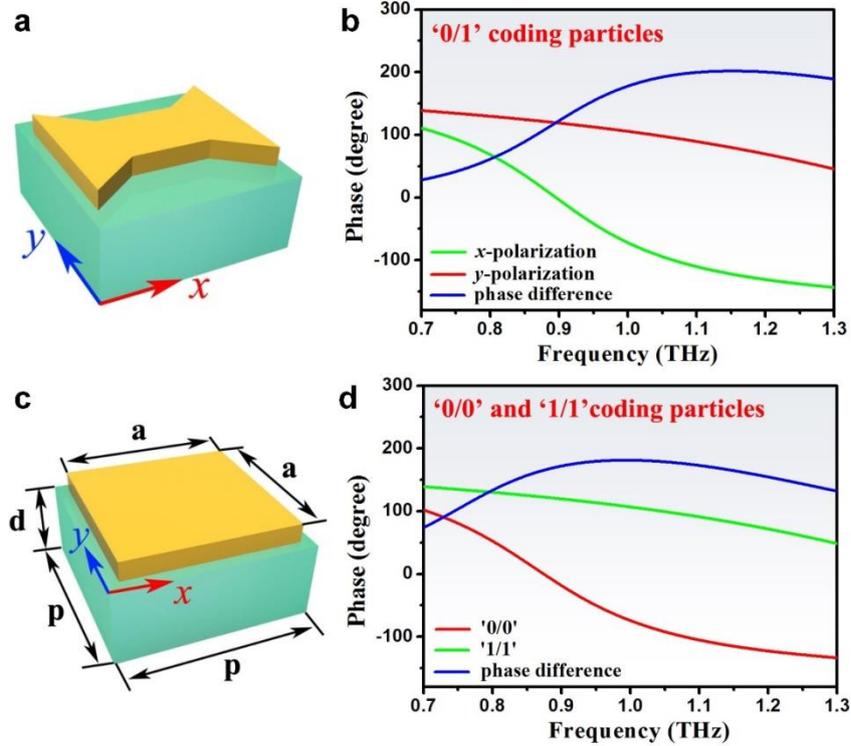

**Supplementary Figure S1 | The structures and reflection responses for the '0/0', '1/1', and '0/1' coding particles.** (**a** and **c**) The structures of the anisotropic unit cell '0/1' and isotropic unit cell '0/0' or '1/1'. (**b**) The reflection phases and corresponding phase difference for the anisotropic unit cell '0/1' under the *x*- and *y*-polarizations. (**d**) The reflection phases and corresponding phase difference for the isotropic unit cell '0/0' and '1/1' under both polarizations.

## Conversion efficiency of the anisotropic coding metamaterial

Despite the versatile functionalities of the polarization-controlled coding metamaterial, the conversion efficiency is also an important factor in real applications. We quantitatively estimate the conversion efficiency of the designed anisotropic coding metamaterial by the ratio of the anomalously deflected/reflected beams to that reflected from a metallic board tilted 24° along the *y*-axis. In this case, the reflected beam points at -48° with respect to the *z*-axis, which is equivalent to the scenario that the same beam is anomalously deflected to -48° by the encoded metasurface. For clear compassion, all the scattering widths in the 2D bistatic scattering patterns have been normalized to the maximum value of the bare metallic case. Supplementary Fig. S2a gives the bi-static scattering pattern of the 1-bit anisotropic coding metasurface, from which we see that the scattering reaches the maximum of 0.456 at -48 ° and 48 ° angle, and the efficiency can be then calculated as 0.456×2=91.2%. Similarly, the conversion efficiencies from the normal incident beam to the anomalously reflected beams are observed from Supplementary Figs. S2d and S3a as 76% and 66%, respectively. Such high efficiency cannot be readily achieved using the single-layered transmitted metasurface because it is difficult to realize the phase variation across $2\pi$ without sacrificing the uniformity of amplitude [1]. We should note that the conversion efficiency defined in this work only represents the proportion of the energy deflected to the direction of 48 °. Different calculation methods may result in different conversion efficiencies [2-4]. Integrating the scattered amplitude over a certain solid angle could produce different results of the conversion efficiency, and may also vary depending on different integration angle of the deflected beam.

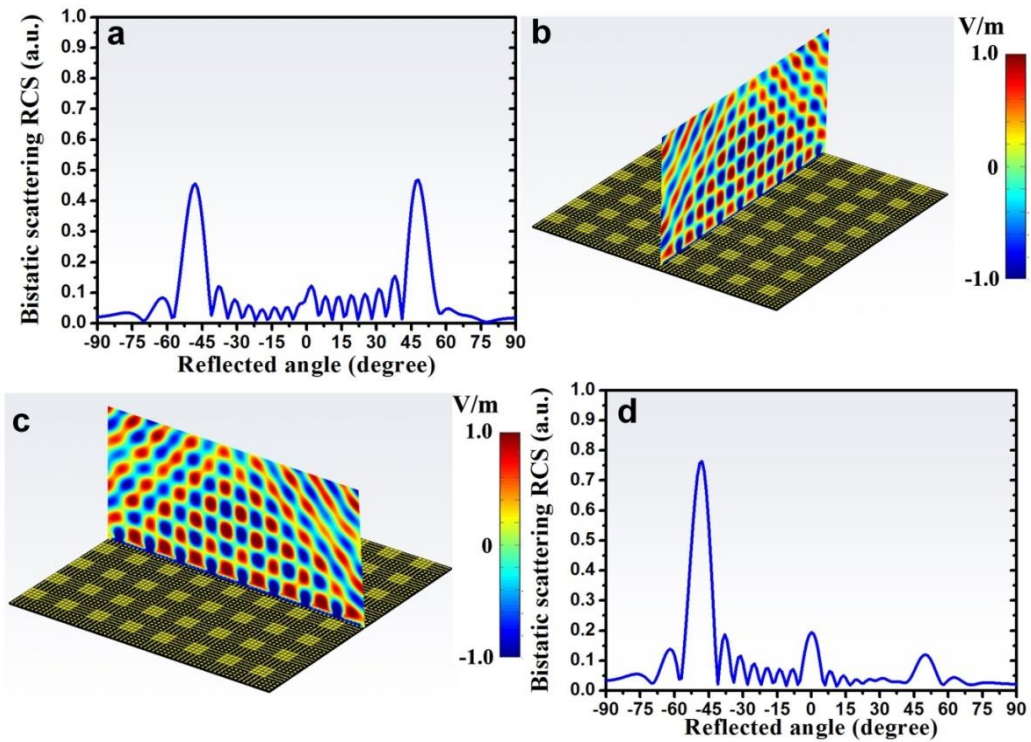

**Supplementary Figure S2 | The simulated results for 1-bit and 2-bit anisotropic coding metasurfaces. (a** and **d)** The 2D far-field scattering patterns for the metasurfaces encoded with coding matrices $\mathbf{M}_1^{1-\text{bit}}$ and $\mathbf{M}_1^{2-\text{bit}}$. (**b** and **c**) The near-electric-field distributions $E_x$ and $E_y$

for the metasurfaces encoded with the coding matrix $\mathbf{M}_1^{1-bit}$ under the *x*- and *y*-polarizations.

**Reflected quarter-wave plate and in-plane beam scanner**

The 2D bi-static scattering pattern for the reflected quarter-wave plate is demonstrated in Supplementary Fig. S3a, where the maximum scattering points to the direction of -30° with amplitude of 0.66, which has been normalized to the maximum value of the beam reflected by metallic board tilted with 15° along the *y*-axis. We notice that the side lobes remain below 0.02 in the whole *x-z* plane. Since the anomalously reflected beam has certain solid angle, we further characterize the axial ratio in a broader angle from -45° to -15° in the *x-z* plane at 1 THz, as shown in Supplementary Fig. S3b. Such excellent performance makes it a high-efficiency and high-directivity device for linear to circular polarization conversion at terahertz frequencies.

In some applications such as a radar system, the beam is usually required to be able to scan from -90° to 90° in a certain plane. This can be easily realized using the proposed 2-bit anisotropic coding metasurface by designing a gradient sequence '00-01-10-11-00-01-10- 11…' for the *x* polarization and its reverse sequence '01-00-11-10-01-00-11-10…' for the *y* polarization, which corresponds to the following coding matrix

$$\mathbf{M}_3^{2-bit}=\begin{pmatrix} 00/01 & 01/00 & 10/11 & 11/10 \\ 00/01 & 01/00 & 10/11 & 11/10 \\ 00/01 & 01/00 & 10/11 & 11/10 \\ 00/01 & 01/00 & 10/11 & 11/10 \end{pmatrix}$$

The size of super unit cell in this case is chosen as 2×2. The final coding pattern containing 32×32 super unit cells is shown in Supplementary Fig. S4a. Supplementary Figures S5a and b give the 3D far-field scattering patterns of the encoded metasurface under the *x*- and *y*-polarizations, respectively, in which two identical anomalously reflected beams are observed at ±48° in the *y-z* plane (see Supplementary Fig.S5c for the 2D plot in the *y-z* plane).

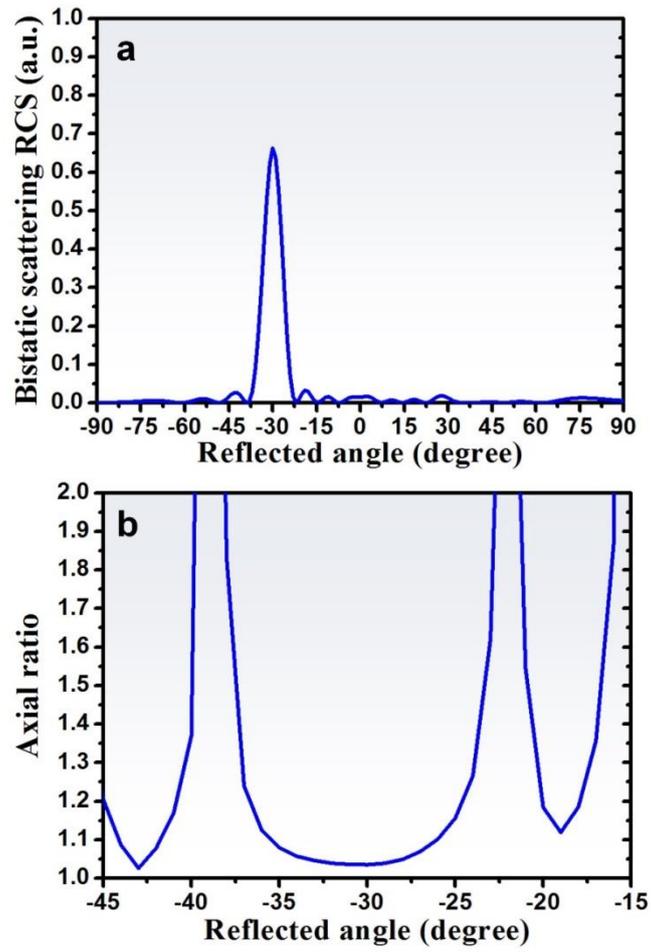

**Supplementary Figure S3 | Simulation results for the reflected quarter-wave plate. (a)** 2D far-field scattering pattern. **(b)** The axial ratio plot from -45 ° to -15 ° in the *x-z* plane at 1 THz.

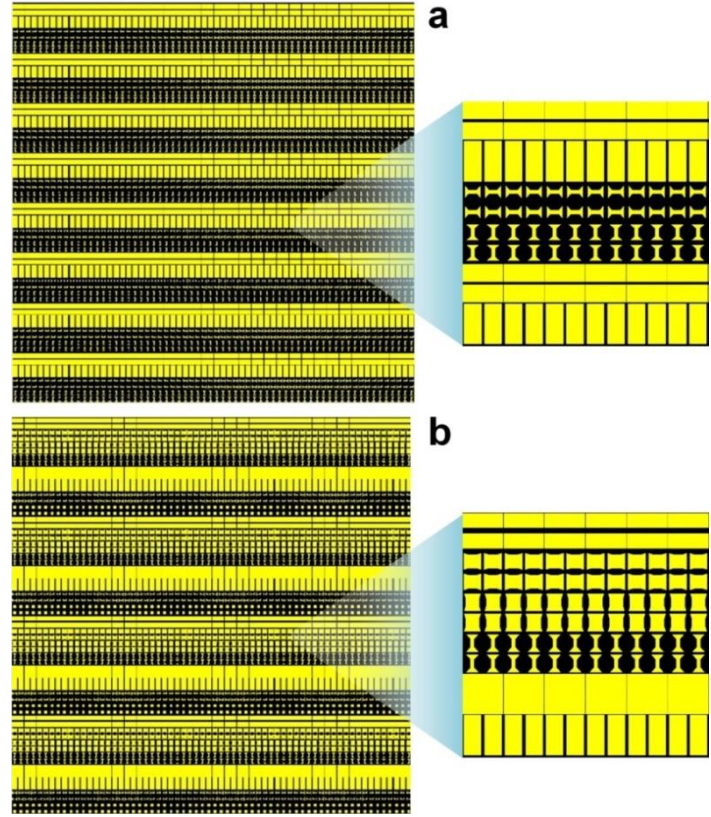

**Supplementary Figure S4 | Coding patterns for 2-bit anisotropic coding metasurfaces and their zoomed views.** (**a**) The pattern with the coding matrix $\mathbf{M}_3^{2-bit}$, which contains 32×32 super unit cells with size 2×2. (**b**) The pattern with the coding matrix $\mathbf{M}_4^{2-bit}$, which contains 32×32 super unit cells with size 2×2.

Similarly, the *y*-polarized terahertz wave will be reflected to the direction of -22° in the *y*-*z* plane if we double the period of the reversed coding sequence as '10-10-01-01-00-00- 11-11-10-10-01-01-00-00-11-11-…', and keeping the coding sequence unchanged for the *x* polarization. Such functionality can be realized by encoding the metasurface with the 8×8 coding matrix

$$\mathbf{M}_4^{2-bit}=\begin{pmatrix} 00/10 & 01/10 & 10/01 & 11/01 & 00/00 & 01/00 & 10/11 & 11/11 \\ 00/10 & 01/10 & 10/01 & 11/01 & 00/00 & 01/00 & 10/11 & 11/11 \\ 00/10 & 01/10 & 10/01 & 11/01 & 00/00 & 01/00 & 10/11 & 11/11 \\ ...... & ...... & ...... & ...... & ...... & ...... & ...... & ...... \end{pmatrix}.$$

The coding pattern containing 32×32 super unit cells (size 2×2) is shown in Supplementary Fig. S4b. Supplementary Figures S5d and e demonstrate the 3D far-field scattering patterns of the encoded metasurface under the *x*- and *y*-polarizations, respectively. In this case, the *y*-polarized terahertz wave is reflected to -22 ° in the *y-z* plane, while the anomalously reflected beam under the *x*-polarization is unaffected. Note that the intensity of -22 ° reflected beam (red curve) has been normalized to the maximum value of the beam reflected by metallic board tilted with 11 ° along the *y*-axis. Again, the difference of the beam intensities of the anomalously reflected beams under x- and y-polarizations may be due to the unpredictable phase responses produced by the interactions between different unit cells. In addition, we note that the eigenmodes under the *x*- and *y*-polarizations are transverse-electric (TE) and transverse-magnetic (TM) modes, respectively, which could also contribute to such discrepancy. Overall, the deflection angle in both cases has strong correlations with the general Snell's law.

**Experiments and measurements**

Supplementary Figure S6c shows the photo of the experimental configuration of the rotational THz-TDS system. In order to minimize the background noise, the received signal was sampled 1000 times at each position of the delay line and, after which an averaged value was calculated as the final measured data. Since the current system does not support 45 ° polarized terahertz wave (with respect to the *y*-axis), the reflected quarter-wave plate design was not measured by this system.

Supplementary Figure S6d displays the photo of the experimental configuration of the theta-to-theta THz-TDS system. The incident wave (marked by blue color) was first reflected by mirror M1 and M2, and then incident to a silicon lens that can transmit half of the energy to the sample and reflect the other half energy back. The anomalously reflected signal (marked by red color), after interacting with the sample, was first reflected by a parabolic mirror and then incident to the sample again. Because the signal passed through the silicon lens twice, the energy of the final received signal was only about a quarter of the incident signal. Supplementary Figs.S6a and b give the optical microscopy images for the 1-bit (encoded with $\mathbf{M}_1^{1-bit}$) and 2-bit (encoded with $\mathbf{M}_1^{2-bit}$) samples, respectively.

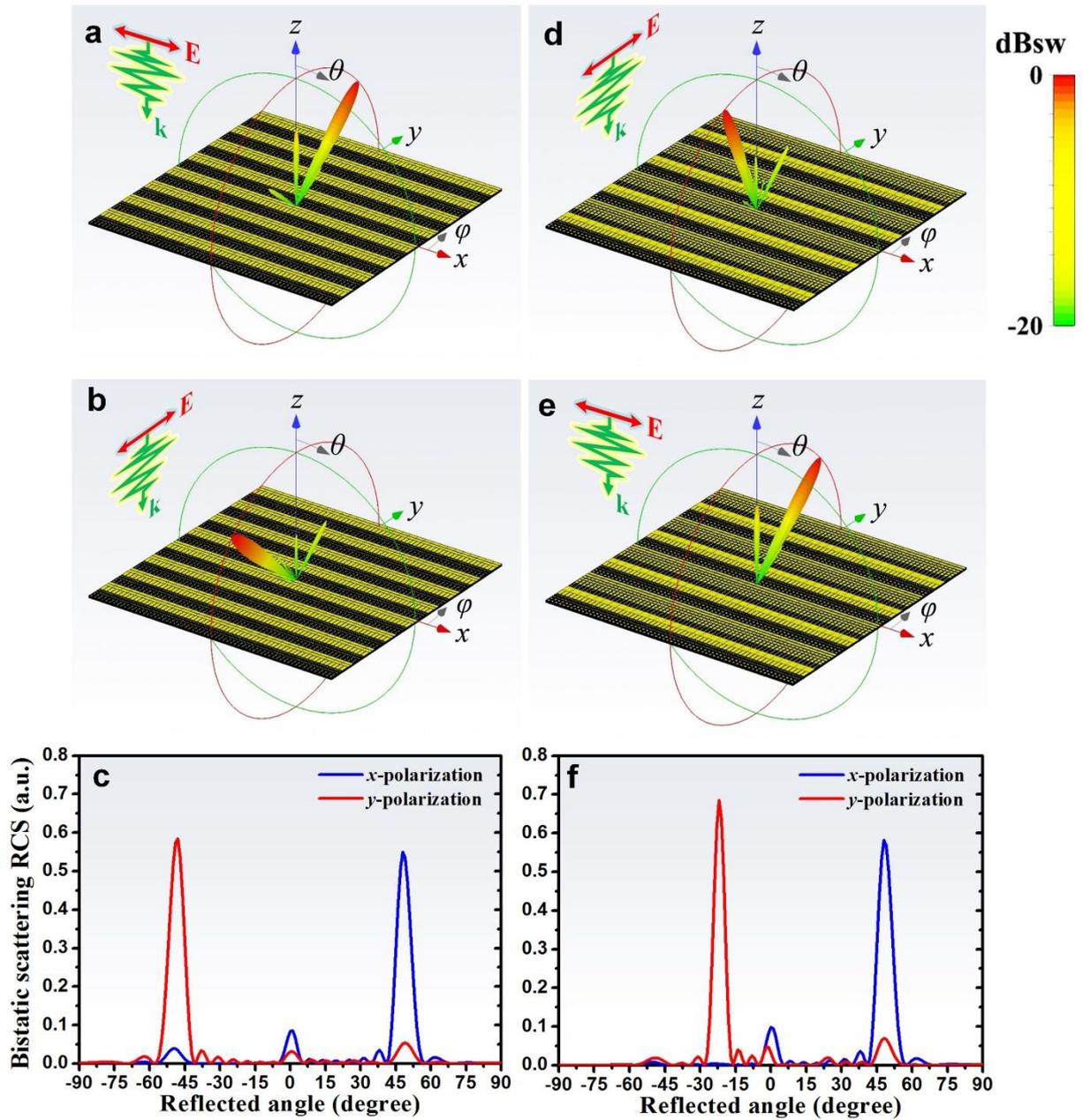

**Supplementary Figure S5 | The simulated 3D and 2D scattering patterns of the in-plane beam scanner.** (**a** and **b**) The 3D far-field scattering patterns of the metasurface encoded with the coding matrix $\mathbf{M}_3^{2-\text{bit}}$ under the *x*- and *y*-polarizations. (**d** and **e**) The 3D far-field scattering patterns of the metasurface encoded with the coding matrix $\mathbf{M}_4^{2-\text{bit}}$ under the *x*- and *y*-polarizations. (**c** and **f**) The 2D far-field scattering patterns of the metasurfaces encoded with coding matrices $\mathbf{M}_3^{2-\text{bit}}$ and $\mathbf{M}_4^{2-\text{bit}}$ under both polarizations in the *y*-*z* plane.

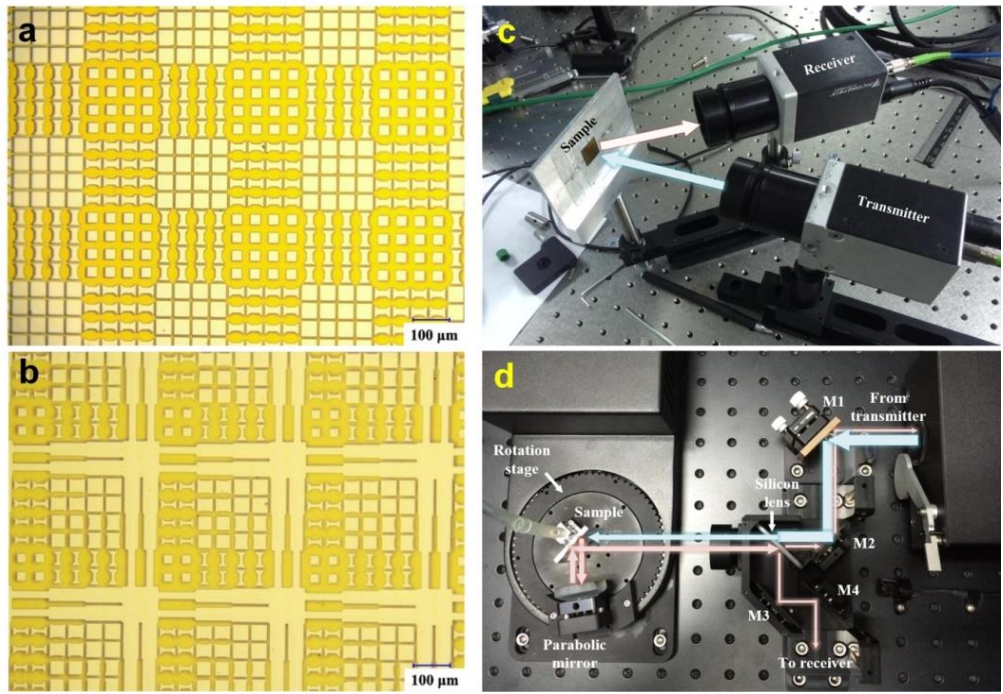

**Supplementary Figure S6 | The photo pictures of fabricated samples and experimental setups. (a** and **b)** The optical microscopy images of the 1-bit (encoded with $\mathbf{M}_1^{1-\text{bit}}$) and the 2-bit (encoded with $\mathbf{M}_1^{2-\text{bit}}$) samples. **(c)** The rotational THz-TDS system. **(d)** The theta-to-theta THz-TDS system.

# References

1. Y. Yao, et al. Broad Electrical Tuning of Graphene-Loaded Plasmonic Antennas. *Nano Lett.* **13**, 1257-1264 (2013).

2. C. Pfeiffer, et al. Efficient Light Bending with Isotropic Metamaterial Huygens' Surfaces *Nano Lett.* **14**, 2491-2497 (2014).

3. M. Kang, T. H. Feng, H. T. Wang, J. S. Li, Wave front engineering from an array of thin aperture antennas. *Opt. Express.* **20**, 15882-15890 (2012).

4. S. L. Sun, et al. High-Efficiency Broadband Anomalous Reflection by Gradient Meta-Surfaces. *Nano Lett.* **12**, 6223-6229 (2012).